\documentclass[english,aps,preprint]{revtex4}
\usepackage[T1]{fontenc}
\usepackage[latin9]{inputenc}
\usepackage{amsmath}
\usepackage{graphicx}

\makeatletter



\usepackage{babel}
\makeatother

\begin{document}

\preprint{}

\title{Physical Reference Frames and Astrometric Measurements of Star Direction
in General Relativity \\I. Stellar Aberration}

\author{Mariateresa Crosta}

\email{crosta@oato.inaf.it}

\affiliation{INAF - Astronomical Observatory of Torino }

\author{Alberto Vecchiato}

\email{vecchiato@oato.inaf.it}

\affiliation{INAF - Astronomical Observatory of Torino}

\begin{abstract}
The high accuracy of modern space astrometry requires the use of General
Relativity to model the propagation of stellar light through the gravitational
field encountered from a source to a given observer inside the Solar
System. In this sense relativistic astrometry is part of fundamental
physics. The general relativistic definition of astrometric measurement
needs an appropriate use of the concept of reference frame, which
should then be linked to the conventions of the IAU Resolutions \citep{2000IAU-res....B1.3},
which fix the celestial coordinate system. A consistent definition
of the astrometric observables in the context of General Relativity
is also essential to find uniquely the stellar coordinates and proper
motion, this being the main physical task of the inverse ray tracing
problem. Aim of this work is to set the level of reciprocal consistency
of two relativistic models, GREM and RAMOD (Gaia, ESA mission), in
order to garantee a physically correct definition of light direction
to a star, an essential item for deducing the star coordinates and
proper motion within the same level of measurement accuracy.
\end{abstract}
\maketitle

\section{introduction}

The correct definition of a physical measurement requires the identification
of an appropriate frame of reference. This applies also to the case
of the determination of position and motion of a star from astrometric
observations made from within our Solar System. Moreover, modern instruments
homed into space-borne astrometric probes like Gaia \citep{2005tdug.conf.....T}
and SIM \citep{2008PASP..120...38U} are targeting accuracy at the
micro-arsecond level, or higher, thus requiring any astrometric measurement
 be modelled in a way that light propagation and detection are both
conceived in a general relativistic framework. One needs, in fact,
to solve the relativistic equations of the null geodesic which describes
the trajectory of a photon emitted by a star and detected by an observer
with an assigned state of motion. The whole process takes place in
a geometrical environment generated by an N-body distribution as could
be that of our Solar System. Essential to the solution of the above
astrometric problem, namely an inverse ray tracing from observational
data, is the identification, as boundary conditions, of the local
observer's line-of-sight defined in a suitable reference frame (see,
e.g. \citep{2003CQGra..20.4695B,2006ApJ...653.1552D,2006CQGra..23.5467D}).

Summarizing from the references quoted above, the astrometric problem
consists in the determination, from a prescribed set of observational
data (hereafter \textit{observables}) of the astrometric parameters
of a star namely its coordinates, parallax, and proper motion. However,
while in classical (non relativistic) astrometry these quantities
are well defined, in General Relativity (GR) they must be interpreted
consistently with the relativistic framework of the model. Similarly,
the parameters describing the attitude and the center-of-mass motion
of the satellite need to be defined consistently with the chosen relativistic
model.

At present, three conceptual frameworks are able to treat the astrometric
problem at the micro-arcsecond level within a relativistic context.

The first model, named GREM (Gaia Relativsitic Model) and described
in \citet{2003AJ....125.1580K}, is an extension of a seminal study
\citet{1992AJ....104..897K} conducted in the framework of the post-Newtonian
(pN) approximation of GR. This model has been formulated according
to a Parametrized Post Newtonian (PPN) scheme accurate to 1~micro-arcsecond.
In this model finite dimensions and angular momentum of the bodies
of the Solar System are included and linked to the motion of the observer
in order to consider the effects of parallax, aberration, and proper
motion. This model is considered as baseline for the Gaia data reduction
\citep{2006oato:avu}. The boundary conditions are fixed by the coordinate
position of the satellite and imposing the value of $c$ to the modulus
of the light direction at past null infinity. The light path is solved
using a matching technique which links the perturbed internal solution
inside the near-zone of the Solar System with the (assumed) flat external
one.

Conceptually similar to the above model is the one developed in \citet{1999PhRvD..60l4002K}.
Using the post-Minkowskian (pM) approximation, Einstein's equations
are solved in the linear regime expressing the perturbated part of
the metric tensor in terms of retarded Lienard-Weichert potentials.
Later, \citet{2002PhRvD..65f4025K} included all the relativistic
effects related to the gravitomagnetic field produced by the traslational
velocity/spin-depedent metric terms.

Both works, in the pN and pM aproaches, rewrite the null geodesic
as function of two independent parameters and solve the light trajectory
as a straight line (Euclidean geometry) plus integrals, containing
the perturbations encountered, from a gravitating source at an arbitrary
distance from an observer located within the Solar System. This allows
one to transform the observed light ray in a suitable coordinate direction
and to read-off the aberrational terms and light deflections effects,
evaluated at the point of observation. The main difference between
the two approximations appears in the computation of the light deflection
contributions: in the pN scheme by the technique of asymptotic matching,
while in the pM one by a semi-analytical integration of the equation
of light propagation from the observer to the source with retarted
time as argument.

The third and last model, RAMOD, is an astrometric model conceived
to solve the inverse ray-tracing problem in a general relativistic
framework not constrained by a priori approximations \citep{2004ApJ...607..580D,2006ApJ...653.1552D}.
It exploits the concept of a curved geometry as a common background
to all steps of its functioning and can be extended to whatever accuracy
and physical requirements \citep{1990recm.book.....D}. RAMOD therefore
is not a just a pN model, contrary to how was referenced in \citep{2006gr.qc....11078T}.
Moreover, the same parametrization of the pN/pM approximations can
be obtained in RAMOD if we limit the model accuracy to the milli-arcsecond
level \citep{2003PhDTh...MRA...C}. The full development to the micro-arcsecond
level imposes to include the $h_{0i}$ metric terms and to take properly
into account the retarded distance effects due to the motion of the
bodies of the Solar System \citep{2006ApJ...653.1552D}. At present,
the RAMOD full solution requires the numerical integration of a set
of coupled non linear differential equations (also called {}``master
equations'') which allows to trace back the light trajectory to the
star initial position and which naturally includes all the effects
due to the curvature of the background geometry. A solution of this
system of differential equations contains all the relativistic perturbations
suffered by the photon along its trajectories due to the intervening
gravitational fields. The boundary conditions fixed by the astrometric
observable as function of an analytical fully relativistic description
of the satellite allows a unique solution for a stellar position and
motion \citep{2003CQGra..20.4695B}.

The first two models, namely the pN and pM ones, though different,
take advantage of a similar {}``language'' that facilitate their
comparison. RAMOD, on the contrary, is formulated in a completely
different way. This makes its comparison with the former two a difficult
task. However, since they are used for the Gaia data reduction with
the purpose to create a catalog of \emph{absolute} positions and proper
motions, any inconsistency in the relativistic model(s) would invalidate
the quality and reliability of the estimates. This alone is sufficient
reason for making a theoretical comparison of the two approaches a
necessity.

In this paper we present the first theoretical comparison, showing
how it is possible to {}``extract'' the aberration terms from the
RAMOD construct.

In section~\ref{sec:The-RAMOD-frames} we review all the building
steps of the RAMOD astrometric set-up. In section~\ref{sec:Single--vs.-Multi-steps}
we compare the procedures used in GREM to those utilized in RAMOD
to define the observables and suggest a possible way to make a comparison
between the quantities of these two formulations via the explicitation
of the aberration part in the RAMOD framework. Section~\ref{sec:aberration-grem}
is devoted to describe the GREM calculations of stellar aberration,
while the following one shows how the same effect can be recovered
in RAMOD. Section~\ref{sec:comments} will finally comment on the
results of the comparison and on some crucial points which have to
be addressed to proceed further with the theoretical comparison of
the two models.

\section{\label{sec:The-RAMOD-frames}The RAMOD frames}

The set-up of any astrometric model implies, primarly, the identification
of the gravitational sources and of the background geometry. Then
one needs to label the space-time points with a coordinate system.
The above steps allow us to fix a reference frame with respect to
which one describes the light trajectory, the motion of the stars
and that of the observer.

The RAMOD framework is based on the weak-field requirement for the
background geometry, which in turn has to be specialized to the particular
case one wants to model. For example, having in mind a Gaia-like mission,
we can assume the Solar System as the only source of gravity, i.e.
a physical system gravitationally bound and weakly relativistic. Then,
only first order terms in the metric perturbation $h$ (or equivalently
in the constant $G$ as in the post-Minkowskian approximation) are
retained. These terms already include all of the possible $(v/c)^{n}$-order
expansions of post-Newtonian approach, but just those up to $(v/c)^{3}$
are needed to reach the micro-arcsecond accuracy required for the
next generation astrometric missions, like e.g. Gaia and SIM.

With these assumptions the background geometry is given by the following
line element\[
\mathrm{d}s^{2}\equiv g_{\alpha\beta}\mathrm{d}x^{\alpha}\mathrm{d}x^{\beta}=\left(\eta_{\alpha\beta}+h_{\alpha\beta}+\mathcal{O}\left(h^{2}\right)\right)\mathrm{d}x^{\alpha}\mathrm{d}x^{\beta},\]
where $\mathcal{O}(h^{2})$ collects all non linear terms in $h$,
the coordinates are $x^{0}=t,x^{1}=x,x^{2}=y,x^{3}=z$, the origin
being fixed at the barycenter of the Solar System, and $\eta_{\alpha\beta}$
is the Minkowskian metric.

For this reason, any comparison between RAMOD and GREM requires that
both use the same metric. In the small curvature limit the metric
components used in RAMOD are \citep{1973grav.book.....M} \begin{equation}
g_{00}=-1+\underset{(2)}{h_{00}}+\mathcal{O}\left(4\right)\qquad g_{0i}=\underset{(3)}{h_{0i}}+\mathcal{O}\left(5\right)\qquad g_{ij}=1+\underset{(2)}{h_{00}}\delta_{ij}+\mathcal{O}\left(4\right),\label{eq:BCRS-metric}\end{equation}
where $\underset{(2)}{h_{00}}=2U/c^{2}$, $\underset{(3)}{h_{0i}}=U^{i}/c^{3}$,
and $U$ and $U^{i}$ are, respectively, the gravitational potential
and the vector potential generated by all the sources inside the Solar
System that can be chosen according the IAU resolution B1.3 \citep{2000IAU-res....B1.3}.
The metric of Eq.~\eqref{eq:BCRS-metric} is also adopted by GREM.
Finally the subscripts indicate the order of $(v/c)$ (e.g. $\underset{(3)}{h_{0i}}\sim\mathcal{O}(3)$
and $\mathcal{O}(n)=\mathcal{O}[(v/c)^{n}]$).

\subsection{The BCRS}

In the near zone of the Solar System and with the metric \eqref{eq:BCRS-metric},
IAU resolutions provide the definition of the Barycentric Celestial
Reference System (BCRS), and of the Satellite Reference System (SRS)
\citep{2000IAU-res....B1.3}. These resolutions, as remarked above,
are based on the pN approximation of GR which is still consistent
with RAMOD, since the perturbation $h_{\alpha\beta}$ to the Minkowskian
metric in \eqref{eq:BCRS-metric}  can be calculated at any desired
order of approximations in $\left(v/c\right)$ inside the Solar System.

In RAMOD (see \citet{2004ApJ...607..580D}) a BCRS is identified requiring
that a smooth family of space-like hypersurfaces exists with equation
$t\left(x,y,z\right)=\mathrm{const}$. The function $t$ can be taken
as a time coordinate. On each of these $t\left(x,y,z\right)=\mathrm{const}$
hypersurfaces one can choose a set of Cartesian-like coordinates centered
at the barycenter of the Solar System (B) and running smoothly as
parameters along curves which point to distant cosmic sources. The
latters are chosen to assure that the system is kinematically non-rotating.
The parameters $x$, $y$, $z$, together with the time coordinate
$t$, provides a basic coordinate representation of the space-time.

Any tensorial quantity will be expressed in terms of coordinate components
relative to coordinate bases induced by the BCRS.

\subsection{The local BCRS}

In RAMOD, at any space-time point there exists a unitary four-vector
$u^{\alpha}$ which is tangent to the world line of a physical observer
at rest with respect to the spatial grid of the BCRS defined as: \begin{equation}
u^{\alpha}=\left(-g_{00}\right)^{-1/2}\delta_{0}^{\alpha}=\left(1+\frac{U}{c^{2}}\right)\mathbf{\partial}_{t}+\mathcal{O}\left(\frac{v^{4}}{c^{4}}\right).\label{eq:ub}\end{equation}

\begin{figure}
\noindent \begin{centering}
\includegraphics[width=0.9\columnwidth]{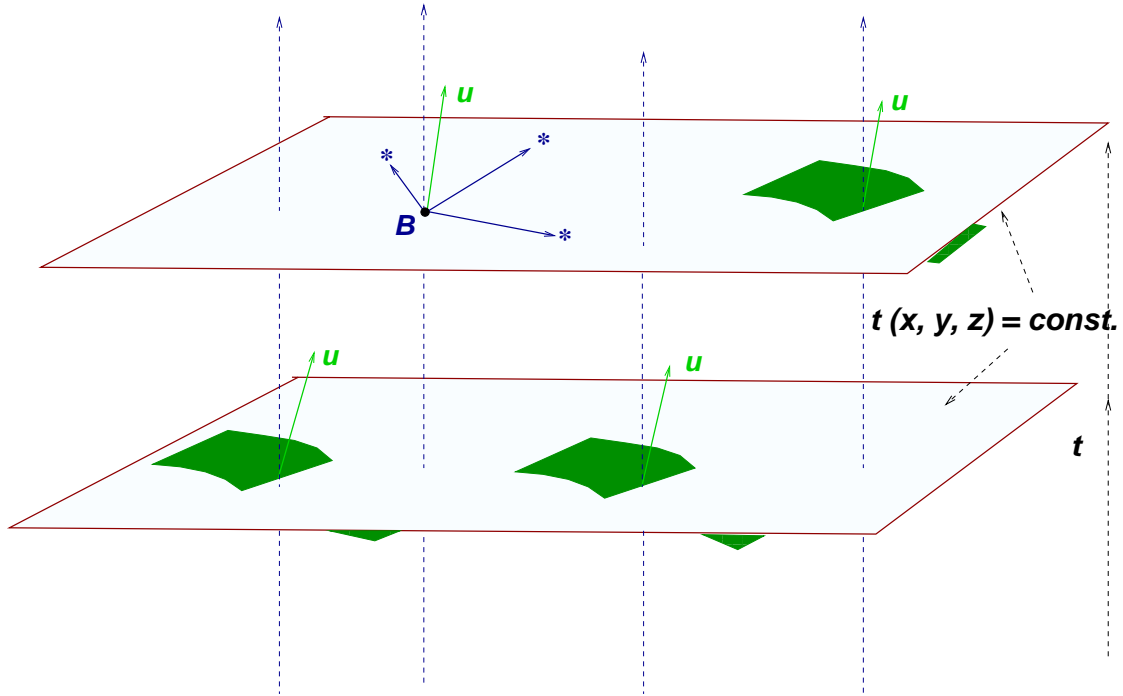} 
\par\end{centering}

\caption{\label{BCRS_local-global} The local observer world line with respect
to the BCRS coordinate system. The spatial axes of the BCRS point
toward distant sources. The dashed lines are the world lines of the
observers at rest with respect to the barycenter (B). The rest-space
(green area) of $\mathbf{u}$ locally deviates from the space-like
hypersurface with equation $t(x,y,z)=\mathrm{const}$ by terms of
the order of a micro-arcsecond.}

\end{figure}

The totality of these four-vectors over the space-time forms a vector
field which is proportional to a time-like and asymptotically Killing
vector field \citep{2004ApJ...607..580D}. The proper time measured
by each of these observers is proportional to the BCRS coordinate
time according to equation \eqref{eq:ub}. To the order of accuracy
required for Gaia, the rest space of $\mathbf{u}$ can be locally
identified by a spatial triad of unitary and orthogonal vectors whose
choice however can only be dictated by specific requirements. A natural
choice is that of pointing to the local coordinate directions chosen
of the BCRS (figure \ref{BCRS_local-global}). %

This frame will be called \emph{local BCRS}; obviously, the local
proper time varies as a function of the gravitational potential at
the observer's position, as can be deduced from equation \eqref{eq:ub}.
In the RAMOD formalism this local BCRS is represented by a tetrad
whose spatial axes (the triad) coincide with the local coordinate
axes, but whose origin is the barycenter of the satellite. At the
$\mathcal{O}\left(h^{2}\right)$, this triad is \citep{2003CQGra..20.4695B}\begin{eqnarray}
\lambda_{\hat{a}}^{\alpha} & = & h_{0a}\delta_{0}^{\alpha}+\left(1-\frac{h_{00}}{2}\right)\delta_{a}^{\alpha}\label{eq:local-bcrs-tetrad}\end{eqnarray}
for $a=1,2,3$.

In RAMOD any physical measurement refers to the local BCRS.

\subsection{The proper reference frame for the satellite}

The proper reference frame of a satellite consists of its rest-space
and a clock which measures the satellite proper time.

The tensorial quantity which expresses a proper reference frame of
a given observer is a \emph{tetrad adapted to that observer}, namely
a set of four unitary mutually orthogonal four-vectors $\mathbf{\mathbf{\lambda}_{\hat{\alpha}}}$
one of which, \textit{i.e.} $\mathbf{\lambda}_{\hat{0}}$, is the
observer's four-velocity while the other $\mathbf{\lambda}_{\hat{a}}$s
form a spatial triad of space-like four-vectors. Mathematically the
tetrad is found as a solution of the following system \citep{1973grav.book.....M}:\begin{equation}
\eta_{ab}=g_{\mu\nu}\lambda_{\hat{a}}^{\mu}\lambda_{\hat{b}}^{\nu}\label{eq:tetrad}\end{equation}
which allows one to interpret a tetrad frame also as an \textit{instantaneous
inertial} reference frame. The solution of \eqref{eq:tetrad}, always
computed w.r.t. the BCRS, is not trivial since it depends on the metric
at each space-time point along the world line of the observer. The
physical measurements made by the observer (satellite) represented
by such a tetrad are obtained by projecting the appropriate tensorial
quantities on the tetrad axes.

The same measurements can also be defined by splitting the space-time
into two \textit{subspaces}. A time-like observer $u'^{\alpha}$ carrying
its laboratory is usually represented as a world tube; in the case
of a non-extended body, the world tube can be restricted to a world
line tracing the history of the observer's barycenter in the given
space-time. At any point $P$ along the world line of $u'^{\alpha}$,
and within a sufficiently small neighborhood, it is possible to split
the space-time into a one-dimensional space and a three-dimensional
one \citep{1990recm.book.....D}, each space being endowed with its
own metric, respectively $U_{\alpha\beta}\left(u'\right)=-u'_{\alpha}u'_{\beta}$
and $P_{\alpha\beta}\left(u'\right)=g_{\alpha\beta}+u'_{\alpha}u'_{\beta}$.
Clearly,

 \begin{equation}
g_{\alpha\beta}=U_{\alpha\beta}\left(u'\right)+P_{\alpha\beta}\left(u'\right).\label{eq:split}\end{equation}

The space with metric $P_{\alpha\beta}\left(u'\right)$ is generated
by lines which stem orthogonally to the world-line of $\mathbf{u'}$
at $P$ and is denoted as the rest-space of the observer $\mathbf{u'}$
at $P$. In this space one measures proper lengths. The space with
metric $U_{\alpha\beta}\left(u'\right)$ is generated by lines which
differ from that of $\mathbf{u'}$ by a riparametrization. In this
space one measures the observer's proper time.

As a consequence of Equation~\eqref{eq:split}, the invariant interval
between two events in space-time can be written as $\mathrm{d}s^{2}=P_{\alpha\beta}\left(u'\right)\,\mathrm{d}x^{\alpha}\mathrm{d}x^{\beta}+U_{\alpha\beta}\left(u'\right)\,\mathrm{d}x^{\alpha}\mathrm{d}x^{\beta}$,
from which we are able to extract the measurements of infinitesimal
spatial distances and times intervals taken by ${\bf u}'$ as, respectively,
$\mathrm{d}L_{u'}=\sqrt{P_{\alpha\beta}\left(u'\right)\mathrm{d}x^{\alpha}\mathrm{d}x^{\beta}}$
and \begin{equation}
\mathrm{d}T_{u'}=-c^{-1}u'_{\alpha}\mathrm{d}x^{\alpha}.\label{eq:dt}\end{equation}

Essentially, the last method is equivalent to the tetrad formalism,
when we do not know the solution of \eqref{eq:tetrad} and we need
to know only the \emph{moduli} of the physical quantities. As far
as RAMOD is concerned, given the metric \eqref{eq:BCRS-metric} and
in the case of a Gaia-like mission, an explicit analitic expression
for a tetrad adapted to the satellite four-velocity exists and can
be found in \citep{2003CQGra..20.4695B}. The spatial axes of this
tetrad are used to model the attitude of the satellite. Moreover,
from  eq.~\eqref{eq:dt}, it is possibile to deduce the IAU trasformations
between the observer's proper time and the barycentric coordinate
time, without using any matching tecnique \citep{2003PhDTh...MRA...C}.
This finally sets the running time on board and completes the definition
of the proper reference frame for the Gaia-like satellite.

\section{\label{sec:Single--vs.-Multi-steps}Single-step vs Multi-step definition
of the observable and a way for the RAMOD vs. GREM comparison}

The classical (non relativistic) approach of astrometry has traditionally
privileged a {}``multi-step'' definition of the observable; i.e.,
the quantities which ultimately enter the {}``final'' catalogue
and are referred to a global inertial reference system, are obtained
taking into account, one by one and independently from each other,
effects such as aberration and parallax.

GREM reproduces in a relativistic framework this approach of classical
astrometry. The BCRS is, for this model, the equivalent of the inertial
reference system of the classical approach, while the final expression
of the star direction in the BCRS is obtained after converting the
observed direction into coordinate ones in several steps which divide
the effects of the aberration, the gravitational deflection, the parallax,
and proper motion \citep{2003AJ....125.1580K}.

In the previous section we have mentioned that RAMOD relies on the
tetrad formalism for the definition of the observable. In general,
the three direction cosines which identify the local line-of-sight
to the observed object are relative to a spatial triad $E_{\hat{a}}$
associated to a given observer $\mathbf{u'}$; the direction cosines
w.r.t. the axes of this triad are defined as:\begin{equation}
\cos\psi_{\hat{a}}=\frac{P(u')_{\alpha\beta}k^{\alpha}E_{\hat{a}}^{\beta}}{\left(P(u')_{\alpha\beta}k^{\alpha}k^{\beta}\right)^{1/2}}\equiv\textbf{e}_{\hat{a}},\label{eq:cos}\end{equation}
where $k^{\alpha}$ is the four-vector tangent to the null geodesic
connecting the star to the observer, and all the quantities are obviously
computed at the event of the observation.

As a consequence of this definition, given the solution of the null
geodesic equation and the motion and the attitude of the observer,
equation~\eqref{eq:cos} expresses a relation between the unknowns,
namely the position and motion of the star, and the observable quantities
which includes all of the above effects mentioned for GREM. In other
words, in RAMOD it is not needed and not natural to disentangle each
single effect, relativistic or not. For this reason any attempt to
make a theoretical comparison between the two models is difficult,
but the way how the observer tetrad was found in RAMOD suggests a
way to overcome this problem.

In \citet{2003CQGra..20.4695B} the attitude frame $E_{\hat{a}}$
was strictly specified for measurements made by a Gaia-like observer.
Let us summarize the main steps.

Given the tetrad $\left\{ \lambda_{\hat{\alpha}}\right\} $ adapted
to the local barycentric observer as defined in (\citet{2003CQGra..20.4695B}
and reference therein) the vectors of the triad $\left\{ \lambda_{\hat{a}}\right\} $
are boosted to the satellite rest frame by means of an instantaneous
Lorentz transformation which depends on the relative spatial velocity
$\nu^{\alpha}$ of the satellite identified by the four-velocity $u'^{\alpha}$
w.r.t. the local BCRS $u^{\alpha}$, and whose Lorentz factor is given
by $\gamma=-u^{\alpha}u'_{\alpha}$ \citep{1992AnnPhys.215..1J}.

The boosted tetrad $\underset{bs}{\lambda^{\alpha}}_{\hat{\alpha}}$
obtained in this way represents, simarly to what is defined for Gaia
in \citep{2004rcn.tech.ll003B,2004PhRvD..69l4001K}, a \textit{\emph{CoMRS}}
(Center-of-Mass Reference System, comoving with the satellite). In
addition to the definition in the cited works one of the axes is \textit{Sun-locked,
i.e.} one axis points toward the Sun at any point of its Lissajous
orbit around L2, in order to deduce the Gaia attitude frame. This
final task is obtained by applying the following rotations to the
\emph{Sun-locked frame}:

\begin{enumerate}
\item by an angle $\omega_{p}t$ about the vector $\underset{bs}{\lambda^{\alpha}}_{\hat{1}}$
which points constantly towards the Sun, where $\omega_{p}$ is the
angular velocity of precession; 
\item by a fixed angle $\alpha$ about the image of the vector $\underset{bs}{\lambda^{\alpha}}_{\hat{2}}$
after the previous rotation; 
\item by an angle $\omega_{r}t$ about the image of the vector $\underset{bs}{\lambda^{\alpha}}_{\hat{1}}$
after the previous two rotations, where $\omega_{r}$ is now the spin
angular velocity. 
\end{enumerate}
The triad resulting from these three steps establishes the satellite
\textit{attitude triad}, given by:\[
E_{\hat{a}}=\mathcal{R}_{1}\left(\omega_{r}t\right)\mathcal{R}_{2}\left(\alpha\right)\mathcal{R}_{1}\left(\omega_{p}t\right)\underset{bs}{\lambda}_{\hat{a}}\qquad a=1,2,3.\]
The final triad $E_{\hat{a}}$ should be the equivalent, in the RAMOD
formalism, to the Satellite Reference System (SRS) \citep{2004rcn.tech.ll003B}
of GREM.

Once this procedure is completed, the final measurements will naturally
entangle in a single result every G R {}``effect''. Therefore, the
natural way to {}``extract'' any of those effects in a separate
formula, is to consider equation~\eqref{eq:cos} and express the
observable as a function of the appropriate tetrad.

\section{\label{sec:aberration-grem}Stellar aberration in GREM}

\begin{figure}
\noindent \begin{centering}
\includegraphics[width=0.9\textwidth]{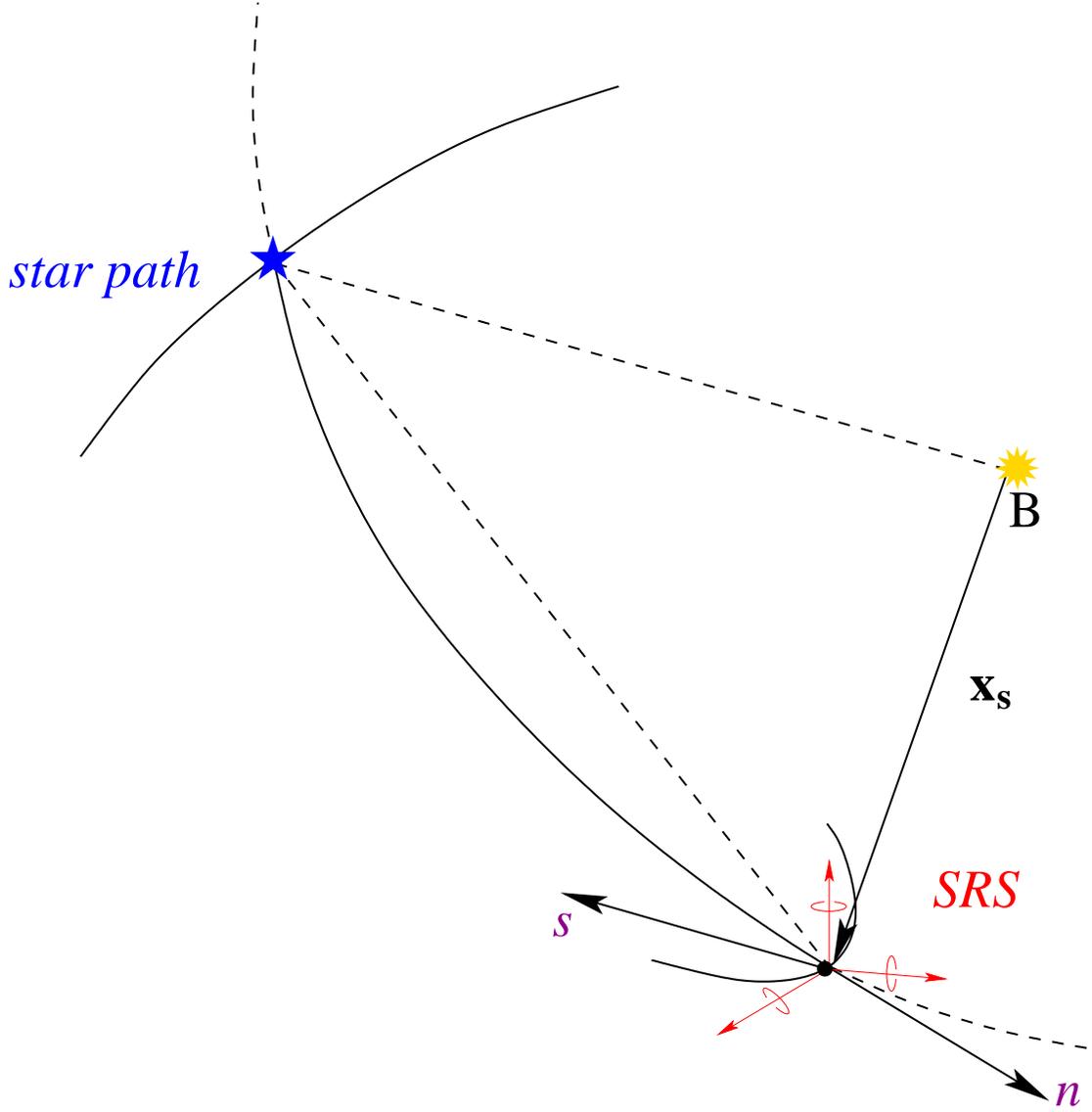} 
\par\end{centering}

\caption{\label{fig:vecs}The vectors representing the light direction in the
pM/pN approches inside the near-zone of the solar system.}

\end{figure}

As well known stellar aberration arises from the motion of the observer
relative to the BCRS origin, assumed to coincide with the center of
mass of the Solar System. In order to account for stellar aberration
in the algorithm for the reduction of the astrometric observations,
the pN/pM approaches \citep{1999PhRvD..60l4002K,2002PhRvD..65f4025K}
transform the observed direction to the source ($\mathbf{s}$) into
the BCRS spatial coordinate direction of the light ray at the point
of observation $\mathbf{x}(t_{\mathrm{o}})\equiv\mathbf{x}_{\mathrm{s}}$
(see figure \ref{fig:vecs}). Now, paraphrasing \citet{2003AJ....125.1580K},
the coordinate direction to the light source at $\mathbf{x}_{\mathrm{s}}$
is defined by the four-vector $p^{\alpha}=(1,p^{i})$, where $p^{i}=c^{-1}dx^{i}/dt$,
$x^{i}$ and $t$ being the BCRS coordinates. But the coordinate components
$p^{i}$ are not a directly observable quantities; the observed vector
towards the light source is the four-vector $s^{\alpha}=(1,s^{i})$,
defined with respect to the local inertial frame of the observer.
In the local frame: \begin{equation}
s^{i}=-\frac{\mathrm{d}\mathcal{X}^{i}}{\mathrm{d}\mathcal{X}^{0}}\label{eq:esse}\end{equation}
where $\mathcal{X}^{\alpha}$ are the coordinates in the CoMRS, then
in order to deduce the spatial direction $p^{i}$ from $s^{i}$ it
is chosen to proceed as follows.

From the property of a null trajectory and taking into account the
metric which defines the BCRS it is

\[
g_{\alpha\beta}p^{\alpha}p^{\beta}=0,\]
namely \begin{eqnarray*}
\left[-1+\frac{\left(1+\gamma_{\mathrm{PPN}}\right)w(t,x)}{c^{2}}-\frac{2w^{2}(t,x)}{c^{4}}\right]+\\
2\delta_{ij}\left(-2\left(1+\gamma\mathrm{_{\mathrm{PPN}}}\right)\frac{w^{i}(t,x)}{c^{3}}\right)p^{j}+\delta_{ij}\left(1+\frac{\left(1+\gamma\mathrm{_{\mathrm{PPN}}}\right)w(t,x)}{c^{2}}\right)p^{i}p^{j} & = & 0\end{eqnarray*}
which gives \begin{equation}
\frac{1}{p}=1+\frac{\left(1+\gamma\mathrm{_{\mathrm{PPN}}}\right)w(t,x)}{c^{2}}-\frac{2\left(1+\gamma\mathrm{_{\mathrm{PPN}}}\right)\delta_{ij}w^{i}(t,x)p^{j}}{c^{3}}+O(c^{-4})\label{eq:mod-p}\end{equation}
where $p=\sqrt{\delta_{ij}p^{i}p^{j}}$ is the Euclidean modulus of
the spatial vector $\mathbf{p}$ and $\gamma\mathrm{_{\mathrm{PPN}}}$
is the PPN parameter.

The infinitesimal transformation $\mathcal{X}^{\alpha}(x^{\beta})$
between CoMRS and BCRS is given by the formula: \begin{equation}
\mathrm{d}\mathcal{X}^{\alpha}=\Lambda_{\beta}^{\alpha}\mathrm{d}x^{\beta}.\label{eq:inftrasf}\end{equation}

From \eqref{eq:inftrasf} the expression of $s^{i}$ as a function
of the spatial components $p^{i}$ is obtained: \begin{equation}
s^{i}=-\frac{\Lambda_{0}^{i}+\Lambda_{j}^{i}p^{j}}{\Lambda_{0}^{0}+\Lambda_{j}^{0}p^{j}}.\label{eq:rel-s-p}\end{equation}

One can explicit formula \eqref{eq:rel-s-p} by following the procedure
reported in \citep{1992AJ....104..897K} and adopting the IAU resolution
B1.3 \citep{2000IAU-res....B1.3}. From the BCRS $\left(ct,x^{i}\right)$
to the CoMRS ($c\mathcal{T},\mathcal{X}^{i}$), the transformation
between the time coordinates reads: \begin{eqnarray}
\mathcal{T} & = & t-c^{-2}[A(t)+\delta_{ij}v^{i}R_{\mathrm{s}}^{j}]\label{eq:tiautras}\\
 &  & +c^{-4}[B+\delta_{ij}B^{i}R_{\mathrm{s}}^{j}+\delta_{im}\delta_{jk}B^{ij}R_{\mathrm{s}}^{m}R_{\mathrm{s}}^{k}+C(t,{\bf x})]+\mathcal{O}\left(c^{-5}\right),\nonumber \end{eqnarray}
and between the spatial coordinates \begin{equation}
\mathcal{X}^{i}=\left[\delta_{j}^{i}+c^{-2}\left(\frac{1}{2}v^{i}v_{j}+qF_{j}^{i}(t)+D_{j}^{i}(t)\right)\right]R_{\mathrm{s}}^{j}+c^{-2}D_{jk}^{i}(t)R_{\mathrm{s}}^{j}R_{\mathrm{s}}^{k}+\mathcal{O}\left(c^{-4}\right).\label{eq:xiautras}\end{equation}
All the functions $A,\, B,\, C,\, D$ are defined in \citet{1992AJ....104..897K}
or in IAU resolutions and \[
R_{\mathrm{s}}^{i}=x^{i}-x_{\mathrm{s}}^{i}\]
are the coordinate displacements with respect to the center of mass
of the satellite $x_{\mathrm{s}}^{i}$ in the BCRS, and finally

\[
v^{i}=\frac{\mathrm{d}x_{\mathrm{s}}^{i}}{\mathrm{d}t}\]
is the coordinate velocity of the center of mass of the satellite
relative to the BCRS.

As reported in \citet{2004PhRvD..69l4001K}, the attitude in GREM
(SRS) is obtained by applying an orthogonal rotation matrix $\mathcal{R}_{i}^{a}$
to $\mathcal{X}^{i}$ in equation~\eqref{eq:xiautras}. At this stage
the role of the SRS is equivalent to that of the $E_{\hat{a}}^{\beta}$s
in eq.~\eqref{eq:cos}.

If one keeps all the terms up to the order of 1~micro-arcsecond,
the observed coordinate direction $s^{i}$, in terms of the unitary
spatial vector $n^{i}=p^{i}/p$, becomes in the CoMRS:\begin{eqnarray}
s^{i} & = & -n^{i}+c^{-1}\left[\mathbf{n}\times\left(\mathbf{v}\times\mathbf{n}\right)\right]^{i}\nonumber \\
 &  & +c^{-2}\left\{ (\mathbf{n}\cdot\mathbf{v})\left[\mathbf{n}\times\left(\mathbf{n}\times\mathbf{v}\right)\right]^{i}+\frac{1}{2}\left[\mathbf{v}\times\left(\mathbf{n}\times\mathbf{v}\right)\right]^{i}\right\} \nonumber \\
 &  & +c^{-3}\left\{ \left[\left(\mathbf{v}\cdot\mathbf{n}\right)^{2}+(1+\gamma_{\mathrm{PPN}})w\left({\bf x_{\mathrm{s}}}\right)\right]\left[\mathbf{n}\times\left(\mathbf{v}\times\mathbf{n}\right)\right]^{i}\right.\nonumber \\
 &  & \left.+\frac{1}{2}\left(\mathbf{n}\cdot\mathbf{v}\right)\left[\mathbf{\mathbf{v}}\times\left(\mathbf{n}\times\mathbf{v}\right)\right]^{i}\right\} +\mathcal{O}\left(\frac{v^{4}}{c^{4}}\right).\label{eq:rel-s-n-grem}\end{eqnarray}

\section{RAMOD aberration in the PM approximation}

Whatever tetrad we consider, the expression of Eq.~\eqref{eq:cos}
for the relativistic observable in the RAMOD model can also be written
as \citep{2006ApJ...653.1552D} \begin{equation}
\mathbf{e}_{\hat{a}}=\frac{\left(\bar{l}_{\left(0\right)}-\nu\right)_{\beta}E_{\hat{a}}^{\beta}}{\gamma\left(1-\nu_{\alpha}\bar{l}_{\left(0\right)}^{\alpha}\right)},\label{eq:rel-obs-1}\end{equation}
where $\nu^{\alpha}$ is the \emph{spatial} four-velocity (also called
as the {}``physical velocity'') of the satellite $\mathbf{u'}$
relative to the \emph{local baricentric observer} $\mathbf{u}$. The
quantity $\bar{l}_{\left(0\right)}^{\alpha}$ was introduced in RAMOD
\citep{2006ApJ...653.1552D} and is a unitary four-vector which represents
the \emph{local line-of-sight} of the photon as seen by $\mathbf{u}$,
i.e. $\bar{l}_{(0)}^{\alpha}=P_{\beta}^{\alpha}(u)k^{\beta}$. 

Finally, $\gamma$ is the Lorentz factor of $u'^{\alpha}$ with respect
to $u^{\alpha}$, that is,\begin{equation}
-u'^{\alpha}u_{\alpha}=\frac{1}{\sqrt{1-\nu^{2}/c^{2}}}\equiv\gamma,\label{eq:gamma}\end{equation}
where $\nu^{2}=\nu^{\alpha}\nu_{\alpha}$.

To retrieve the aberration effect given by the motion of the satellite
with respect to the BCRS in RAMOD, one needs to specialize Eq.~\eqref{eq:rel-obs-1}
to the case of a tetrad $\left\{ \tilde{\lambda}_{\hat{\alpha}}\right\} $
adapted to the center of mass of the satellite assumed with no attitude
parameters. In this case, in fact, the observation equation will give
a relation between the {}``aberrated'' direction represented by
the direction cosines $\cos\psi_{a}$ as measured by the satellite
and the {}``aberration-free'' direction given by the quantity $\bar{l}_{\left(0\right)}^{\alpha}$
referred to the local BCRS frame $\lambda_{\hat{a}}^{\alpha}$. The
vectors of the triad $\left\{ \tilde{\lambda}_{\hat{a}}\right\} $
differ from the local BCRS's $\left\{ \lambda_{\hat{a}}\right\} $
for a boost transformation with four-velocity $u'^{\alpha}$. This
means that it can be derived from Eq.~\eqref{eq:local-bcrs-tetrad}
using the relation \citep{1992AnnPhys.215..1J}\begin{equation}
\tilde{\lambda}_{\hat{a}}^{\alpha}=P\left(u'\right)_{\sigma}^{\alpha}\left[\lambda_{\hat{a}}^{\sigma}-\frac{\gamma}{\gamma+1}\nu^{\sigma}\left(\nu^{\rho}\lambda_{\rho\hat{a}}\right)\right],\label{eq:boost-transf}\end{equation}
where $u'^{\alpha}$ and $\nu^{\alpha}$ are the above mentioned four-velocity
of the satellite and its physical velocity relative to the local BCRS
respectively, and $P\left(u'\right)_{\sigma}^{\alpha}=\delta_{\sigma}^{\alpha}+u'^{\alpha}u'_{\sigma}$.

From \citet{2006ApJ...653.1552D} and \citet{2003CQGra..20.4695B}
it is\begin{eqnarray}
u'^{\alpha} & = & \left(1+\frac{U}{c^{2}}+\frac{1}{2}\frac{v^{2}}{c^{2}}\right)\left(\delta_{0}^{\alpha}+\frac{v^{i}}{c}\delta_{i}^{\alpha}\right)+\mathcal{O}\left(\frac{v^{4}}{c^{4}}\right)\label{eq:4-vel-obs}\end{eqnarray}
where $v^{i}$ is the coordinate velocity of the satellite, as stated
in the previous section. Now, being \citep{2006ApJ...653.1552D}\begin{equation}
\nu^{\alpha}=\frac{1}{\gamma}\left(u'^{\alpha}-\gamma u^{\alpha}\right),\label{eq:rel-u-phys-vel}\end{equation}
one deduces that $\nu^{0}\sim\mathcal{O}\left(v^{4}/c^{4}\right)$
and\begin{eqnarray}
\nu^{i} & = & \left(1+\frac{U}{c^{2}}\right)\frac{v^{i}}{c}+\mathcal{O}\left(\frac{v^{4}}{c^{4}}\right).\label{eq:phys-vel-obs}\end{eqnarray}

Expanding Eq.~\eqref{eq:boost-transf}  with relations \eqref{eq:gamma}
and \eqref{eq:rel-u-phys-vel} one gets

\begin{eqnarray}
\tilde{\lambda}_{\hat{a}}^{\alpha} & = & \lambda_{\hat{a}}^{\alpha}+u'^{\alpha}\left(u'|\lambda_{\hat{a}}\right)-\left(\frac{1}{2}+\frac{1}{8}\frac{v^{2}}{c^{2}}\right)\nu^{\alpha}\left(\nu|\lambda_{\hat{a}}\right)\nonumber \\
 & - & u'^{\alpha}\left(\frac{1}{2}+\frac{1}{8}\frac{v^{2}}{c^{2}}\right)\left(u'|\nu\right)\left(\nu|\lambda_{\hat{a}}\right)+\mathcal{O}\left(\frac{v^{4}}{c^{4}}\right)\label{eq:bst-tetrad-1}\end{eqnarray}
where the notation $\left(\cdot|\cdot\right)$ represents the scalar
product, so, e.g., $\left(u'|\lambda_{\hat{a}}\right)=g_{\alpha\beta}u'^{\alpha}\lambda_{\hat{a}}^{\beta}$.

Then, using Eqs.~\eqref{eq:local-bcrs-tetrad} \eqref{eq:4-vel-obs}
and \eqref{eq:phys-vel-obs} and expanding the scalar products to
the right order we obtain\begin{eqnarray}
\left(u'|\lambda_{\hat{a}}\right) & = & \left(1+\frac{2U}{c^{2}}+\frac{1}{2}\frac{v^{2}}{c^{2}}\right)\frac{v^{a}}{c}+\mathcal{O}\left(\frac{v^{4}}{c^{4}}\right)\label{eq:udotl}\\
\left(\nu|\lambda_{\hat{a}}\right) & = & \left(1+\frac{2U}{c^{2}}\right)\frac{v^{a}}{c}+\mathcal{O}\left(\frac{v^{4}}{c^{4}}\right)\label{eq:nudotl}\\
\left(u'|\nu\right) & = & \frac{v^{2}}{c^{2}}+\mathcal{O}\left(\frac{v^{4}}{c^{4}}\right)\label{eq:udotnu}\end{eqnarray}
so that the expression for the boosted tetrad finally becomes\begin{equation}
\tilde{\lambda}_{\hat{a}}^{\alpha}=\lambda_{\hat{a}}^{\alpha}+\left(1+\frac{3U}{c^{2}}+\frac{1}{2}\frac{v^{2}}{c^{2}}\right)\delta_{0}^{\alpha}\frac{v^{a}}{c}+\frac{1}{2}\frac{v^{i}}{c}\delta_{i}^{\alpha}\frac{v^{a}}{c}+\mathcal{O}\left(\frac{v^{4}}{c^{4}}\right).\label{eq:bstd-tetrad-2}\end{equation}

\subsection{The $\left(v/c\right)^{3}$ expansion of the relativistic observable}

Given Eq.~\eqref{eq:bstd-tetrad-2} one can consistently recast Eq.~\eqref{eq:rel-obs-1}
as\begin{equation}
\mathbf{\tilde{e}}_{\hat{a}}=\frac{\left(\bar{l}-\nu\right)_{\beta}\lambda_{\hat{a}}^{\beta}}{\gamma\left(1-\nu_{\alpha}\bar{l}^{\alpha}\right)}+\frac{\left(\bar{l}-\nu\right)_{\beta}\delta_{0}^{\beta}}{\gamma\left(1-\nu_{\alpha}\bar{l}^{\alpha}\right)}\left(1+\frac{3U}{c^{2}}+\frac{1}{2}\frac{v^{2}}{c^{2}}\right)\frac{v^{a}}{c}+\frac{1}{2}\frac{\left(\bar{l}-\nu\right)_{\beta}\frac{v^{i}}{c}\delta_{i}^{\beta}}{\gamma\left(1-\nu_{\alpha}\bar{l}^{\alpha}\right)}\frac{v^{a}}{c}+\mathcal{O}\left(\frac{v^{4}}{c^{4}}\right)\label{eq:rel-obs-2}\end{equation}
where $\mathbf{\tilde{e}}_{\hat{a}}$ are the cosines related to the
tetrad $\tilde{\lambda}_{\hat{a}}^{\alpha}$ which, as said, does
not contain the attitude parameters. Here and in the rest of the section,
we replace the symbol $\bar{l}_{\left(0\right)}^{\alpha}$ with $\bar{l}^{\alpha}$
to ease the notation.

After long calculations, the first term on the right-hand-side of
this formula can be written as \begin{eqnarray}
\frac{\left(\bar{l}-\nu\right)_{\beta}\lambda_{\hat{a}}^{\beta}}{\gamma\left(1-\nu_{\alpha}\bar{l}^{\alpha}\right)} & = & \bar{l}^{a}+\frac{1}{c}\left[-v^{a}+\left(\delta_{ij}v^{i}\bar{l}^{j}\right)\bar{l}^{a}\right]+\nonumber \\
 &  & \frac{1}{c^{2}}\left\{ U\bar{l}^{a}-\left(\delta_{ij}v^{i}\bar{l}^{j}\right)v^{a}+\left[\left(\delta_{ij}v^{i}\bar{l}^{j}\right)^{2}-\frac{1}{2}v^{2}\right]\bar{l}^{a}\right\} +\nonumber \\
 &  & \frac{1}{c^{3}}\left\{ -2Uv^{a}-\left[\left(\delta_{ij}v^{i}\bar{l}^{j}\right)^{2}-\frac{1}{2}v^{2}\right]v^{a}+\right.\nonumber \\
 &  & \phantom{\frac{1}{c^{3}}\{}\left.\bar{l}^{a}\left[3U\left(\delta_{ij}v^{i}\bar{l}^{j}\right)+\left(\delta_{ij}v^{i}\bar{l}^{j}\right)^{3}-\frac{1}{2}v^{2}\left(\delta_{ij}v^{i}\bar{l}^{j}\right)+U\left(\delta_{ij}v^{i}\bar{l}^{j}\right)\right]\right\} \nonumber \\
 &  & +\mathcal{O}\left(\frac{v^{4}}{c^{4}}\right),\label{eq:first-term}\end{eqnarray}
the second term is zero since both $\bar{l}_{0}$and $\nu_{0}$ are
zero, while the third one becomes \begin{eqnarray}
\frac{1}{2}\frac{\left(\bar{l}-\nu\right)_{i}\left(v^{i}/c\right)}{\gamma\left(1-\nu_{\alpha}\bar{l}^{\alpha}\right)}\frac{v^{a}}{c} & = & \frac{1}{2}\left[\delta_{ij}\bar{l}^{i}\frac{v^{j}}{c}+\left(\delta_{ij}\frac{v^{i}}{c}\bar{l}^{j}\right)^{2}-\frac{v^{2}}{c^{2}}\right]\frac{v^{a}}{c}+\mathcal{O}\left(\frac{v^{4}}{c^{4}}\right).\label{eq:third-term}\end{eqnarray}

Finally, collecting all terms:\begin{eqnarray}
\mathbf{\tilde{e}}_{\hat{a}} & = & \bar{l}^{a}+\frac{1}{c}\left[-v^{a}+\left(\delta_{ij}v^{i}\bar{l}^{j}\right)\bar{l}^{a}\right]+\nonumber \\
 &  & \frac{1}{c^{2}}\left\{ U\bar{l}^{a}-\frac{1}{2}\left(\delta_{ij}v^{i}\bar{l}^{j}\right)v^{a}+\left[\left(\delta_{ij}v^{i}\bar{l}^{j}\right)^{2}-\frac{1}{2}v^{2}\right]\bar{l}^{a}\right\} +\nonumber \\
 &  & \frac{1}{c^{3}}\left\{ -2Uv^{a}-\frac{1}{2}\left(\delta_{ij}v^{i}\bar{l}^{j}\right)^{2}v^{a}+\right.\nonumber \\
 &  & \phantom{\frac{1}{c^{3}}\{}\left.\bar{l}^{a}\left[3U\left(\delta_{ij}v^{i}\bar{l}^{j}\right)+\left(\delta_{ij}v^{i}\bar{l}^{j}\right)^{3}-\frac{1}{2}v^{2}\left(\delta_{ij}v^{i}\bar{l}^{j}\right)+U\left(\delta_{ij}v^{i}\bar{l}^{j}\right)\right]\right\} +\mathcal{O}\left(\frac{v^{4}}{c^{4}}\right).\label{eq:all-terms}\end{eqnarray}
At a first glance, the last expression shows differences in terms
up to the $\left(v/c\right)^{2}$ order (note in particular the appearance
of the term $U\bar{l}^{a}$) and of the $\left(v/c\right)^{3}$ order
which cannot allow to straightforwardly compare, as expected, the
above expression to the GREM vectorial one of Eq.~\eqref{eq:rel-s-n-grem}.

\subsection{Comparison with the GREM model}

The expression~\eqref{eq:all-terms} relates the observed direction
cosines with $\bar{l}^{\alpha}$. The equivalent relation for the
GREM observable is equation~\eqref{eq:rel-s-n-grem} where the aberration
is expressed in terms of a vector \textbf{$\mathbf{n}$}. To compare
formula~\eqref{eq:all-terms} with GREM's formula~\eqref{eq:rel-s-n-grem}
we need to find a relationship between \textbf{$\mathbf{n}$} and
$\bar{l}^{\alpha}$. To this purpose we need to reduce $\bar{l}^{\alpha}$
to its coordinate euclidean expression

In GREM $\mathbf{n}$ represents the {}``aberration-free'' coordinate
line of sight of the observed star at the position of the satellite
momentarily at rest. In RAMOD, as said, $\bar{l}^{\alpha}$ represents
the normalized \emph{local} \emph{line-of-sight} of the observed star
\emph{as} \emph{seen} by the local barycentric observer $\mathbf{u}$.
In other words, $\bar{l}^{\alpha}$ is a four-vector which fixes the
line-of-sight of an object with respect to the local BCRS.

Do $\mathbf{n}$ and $\bar{l}^{\alpha}$ have a similar role in the
two approaches? From the physical point of view they have the same
meaning, as the \emph{observed} {}``aberration free'' direction
to the star. Let us start from the definition of $\mathbf{n}$ in
GREM:\[
n^{i}=\frac{p^{i}}{p},\]
where \emph{$p^{i}=c^{-1}\mathrm{d}x^{i}/\mathrm{d}t$} and $p$ is
the Euclidean norm of $p^{i}$, so that $p^{-1}\simeq\left(1+h_{00}+h_{0i}p^{i}\right)+\mathcal{O}\left(h^{2}\right)$,
as equation~\eqref{eq:mod-p} shows. This means that \begin{equation}
n^{i}=p^{i}\left(1+h_{00}+h_{0i}p^{i}\right)+\mathcal{O}\left(h^{2}\right).\label{eq:expr-ni}\end{equation}
On the other hand, using the definition of $\bar{l}^{\alpha}$ in
\citet{2006ApJ...653.1552D} it can be easily shown that its spatial
components are\[
\bar{l}^{i}=-\frac{k^{i}}{u_{\alpha}k^{\alpha}}=-\frac{k^{i}}{u^{0}k^{0}\left(-1+h_{00}+h_{0i}\frac{k^{i}}{k^{0}}\right)},\]
and, from $u^{0}=\left(-g_{00}\right)^{-1/2}$ and $k^{i}/k^{0}=c^{-1}\mathrm{d}x^{i}/\mathrm{d}t\equiv p^{i}$,
it results\begin{eqnarray}
\bar{l}^{i} & = & p^{i}\left(-g_{00}\right)^{1/2}\left(1-h_{00}-h_{0i}p^{i}\right)^{-1}\nonumber \\
 & = & p^{i}\left(1+\frac{1}{2}h_{00}+h_{0i}p^{i}\right)+\mathcal{O}\left(h^{2}\right).\label{eq:expr-li}\end{eqnarray}
Finally, from equations~\eqref{eq:expr-ni} and \eqref{eq:expr-li}
one has\begin{eqnarray}
\bar{l}^{i} & = & n^{i}\left(1-\frac{U}{c^{2}}\right)+\mathcal{O}\left(\frac{v^{4}}{c^{4}}\right)\label{eq:rel-li-ni}\end{eqnarray}
namely, the spatial light direction, expressed in terms of its Euclidean
counterpart at the satellite location in the gravitational field of
the solar system. Worth noticing is that no terms of the order of
$\mathcal{O}[(v/c)^{3}]$ appear in \eqref{eq:rel-li-ni}.

Combining Eq.~\eqref{eq:all-terms} with \eqref{eq:rel-li-ni} and
setting $\left(\delta_{ij}v^{i}n^{j}\right)\equiv\mathbf{v}\cdot\mathbf{n}$
to ease the notation, we obtained\begin{eqnarray}
\mathbf{\tilde{e}}_{\hat{a}} & = & n^{a}+\frac{1}{c}\left[-v^{a}+\left(\mathbf{v}\cdot\mathbf{n}\right)n^{a}\right]+\frac{1}{c^{2}}\left\{ -\frac{1}{2}\left(\mathbf{v}\cdot\mathbf{n}\right)v^{a}+\left[\left(\mathbf{v}\cdot\mathbf{n}\right)^{2}-\frac{1}{2}v^{2}\right]n^{a}\right\} +\nonumber \\
 &  & \frac{1}{c^{3}}\left\{ -2Uv^{a}-\frac{1}{2}\left(\mathbf{v}\cdot\mathbf{n}\right)^{2}v^{a}+\left(\mathbf{v}\cdot\mathbf{n}\right)n^{a}\left[2U+\left(\mathbf{v}\cdot\mathbf{n}\right)^{2}-\frac{1}{2}v^{2}\right]\right\} +\mathcal{O}\left(\frac{v^{4}}{c^{4}}\right).\label{eq:cospsia-1}\end{eqnarray}

In this way the right-hand side of the aberration expression of RAMOD
is rewritten with the GREM quantities at the $\left(v/c\right)^{3}$
order. The same operation can be done for the left-hand side using
the definition of the projection operator and the tetrad property
$\lambda_{\alpha}^{\hat{\mu}}\lambda_{\hat{\mu}\beta}=g_{\alpha\beta}$:
\begin{equation}
\mathbf{\tilde{e}}_{\hat{a}}\equiv\frac{P(u)_{\alpha\beta}k^{\alpha}\tilde{\lambda}_{\hat{a}}^{\beta}}{\left(P(u)_{\alpha\beta}k^{\alpha}k^{\beta}\right)^{1/2}}=\frac{k^{\alpha}\tilde{\lambda}_{\alpha}^{\hat{a}}}{|g_{\alpha\beta}u^{\alpha}k^{\beta}|}=-\frac{k^{\alpha}\tilde{\lambda}_{\alpha}^{\hat{a}}}{g_{\alpha\beta}\tilde{\lambda}_{\hat{0}}^{\alpha}k^{\beta}}=\frac{k^{\alpha}\tilde{\lambda}_{\alpha}^{\hat{a}}}{k^{\beta}\tilde{\lambda}_{\beta}^{\hat{0}}}=\frac{\mathrm{d}\tilde{x}^{\hat{a}}}{\mathrm{d}\tilde{x}^{\hat{0}}}.\label{eq:cos-s}\end{equation}

Is there a relation between the direction cosines of the above equation
with the spatial components of the observed vector $s^{i}$ in GREM?
The crucial point stands on the definition of the coordinates system.
The tetrad components of the light ray can be directly associated
to CoMRS coordinates (as done in \citet{2004PhRvD..69l4001K}) if
the boosted local BCRS tetrad coordinates $\tilde{x}^{\hat{\alpha}}$
are equivalent to the CoMRS ones $\mathcal{X}^{\alpha}$. This is
true only locally, i.e. in a sufficiently small neighborhood (since
the tetrad are not in general olonomous) and if the origins of the
two reference systems concide. So, from \eqref{eq:esse}, if one could
state that

\[
\frac{\mathrm{d}\tilde{x}^{\hat{a}}}{\mathrm{d}\tilde{x}^{\hat{0}}}=\frac{\mathrm{d}\mathcal{X}^{a}}{\mathrm{d}\mathcal{X}^{0}}\equiv-s^{a},\]
it would follow \begin{equation}
\mathbf{\tilde{e}}_{\hat{a}}=-s^{a}.\label{eq:cos_esse}\end{equation}

In RAMOD, at the milli-arcsecond level, the rest space of the local
baricentric observer coincides globally with the spatial hypersurfaces
which foliate the space-time and define the BCRS \citep{2004ApJ...607..580D}.
At micro-arcsecond accuracy, instead, the vorticity cannot be neglected
and the geometry is affected by non-diagonal terms of the metric hence
the $t=\mathrm{constant}$ hypersurfaces do not coincide with the
rest-space of the local barycentric observer \citep{2006ApJ...653.1552D}.
Then, to be consistent we can only define at each point of observation
a spatial direction measured by the local barycentric observer and
then associate it to the satellite measurements via the direction
cosines relative to the boosted attitude frame. As far as GREM is
concerned, the euclidean geometry admits a parallel transport which
does not feel the curvature, allowing to define the same vector in
any point of the space. 

Then, equation~\eqref{eq:cos_esse} has only local validity and \eqref{eq:cospsia-1}can
be written as\begin{eqnarray}
-s^{a} & = & n^{a}+\frac{1}{c}\left[-v^{a}+\left(\mathbf{v}\cdot\mathbf{n}\right)n^{a}\right]+\frac{1}{c^{2}}\left\{ -\frac{1}{2}\left(\mathbf{v}\cdot\mathbf{n}\right)v^{a}+\left[\left(\mathbf{v}\cdot\mathbf{n}\right)^{2}-\frac{1}{2}v^{2}\right]n^{a}\right\} +\nonumber \\
 &  & \frac{1}{c^{3}}\left\{ -2Uv^{a}-\frac{1}{2}\left(\mathbf{v}\cdot\mathbf{n}\right)^{2}v^{a}+\left(\mathbf{v}\cdot\mathbf{n}\right)n^{a}\left[2U+\left(\mathbf{v}\cdot\mathbf{n}\right)^{2}-\frac{1}{2}v^{2}\right]\right\} +\mathcal{O}\left(\frac{v^{4}}{c^{4}}\right).\label{eq:sa-1}\end{eqnarray}
Considering that $\mathbf{n}\cdot\mathbf{n}=1$ and $v^{2}=\delta_{ij}v^{i}v^{j}\equiv\mathbf{v}\cdot\mathbf{v}$,
the previous equation becomes\begin{eqnarray}
s^{a} & = & -n^{a}+\frac{1}{c}\left[v^{a}\left(\mathbf{n}\cdot\mathbf{n}\right)-n^{a}\left(\mathbf{v}\cdot\mathbf{n}\right)\right]+\nonumber \\
 &  & \frac{1}{c^{2}}\left\{ \left(\mathbf{v}\cdot\mathbf{n}\right)\left[v^{a}\left(\mathbf{n}\cdot\mathbf{n}\right)-n^{a}\left(\mathbf{v}\cdot\mathbf{n}\right)\right]+\frac{1}{2}\left[n^{a}\left(\mathbf{v}\cdot\mathbf{v}\right)-v^{a}\left(\mathbf{v}\cdot\mathbf{n}\right)\right]\right\} +\nonumber \\
 &  & \frac{1}{c^{3}}\left\{ 2U\left[v^{a}\left(\mathbf{n}\cdot\mathbf{n}\right)-n^{a}\left(\mathbf{v}\cdot\mathbf{n}\right)\right]+\left(\mathbf{v}\cdot\mathbf{n}\right)^{2}\left[v^{a}\left(\mathbf{n}\cdot\mathbf{n}\right)-n^{a}\left(\mathbf{v}\cdot\mathbf{n}\right)\right]+\right.\nonumber \\
 &  & \phantom{\frac{1}{c^{3}}\{}\left.\frac{1}{2}\left(\mathbf{v}\cdot\mathbf{n}\right)\left[n^{a}\left(\mathbf{v}\cdot\mathbf{v}\right)-v^{a}\left(\mathbf{v}\cdot\mathbf{n}\right)\right]\right\} +\mathcal{O}\left(\frac{v^{4}}{c^{4}}\right).\label{eq:sa-2}\end{eqnarray}
Finally, from the relation $\mathbf{a}\times\left(\mathbf{b}\times\mathbf{c}\right)=\mathbf{b}\left(\mathbf{a}\cdot\mathbf{c}\right)-\mathbf{c}\left(\mathbf{a}\cdot\mathbf{b}\right)$
it is \begin{eqnarray}
s^{a} & = & -n^{a}+\frac{1}{c}\left[\mathbf{n}\times\left(\mathbf{v}\times\mathbf{n}\right)\right]^{a}+\frac{1}{c^{2}}\left\{ \left(\mathbf{v}\cdot\mathbf{n}\right)\left[\mathbf{n}\times\left(\mathbf{v}\times\mathbf{n}\right)\right]^{a}+\frac{1}{2}\left[\mathbf{v}\times\left(\mathbf{n}\times\mathbf{v}\right)\right]^{a}\right\} +\nonumber \\
 &  & \frac{1}{c^{3}}\left\{ \left[\left(\mathbf{v}\cdot\mathbf{n}\right)^{2}+2U\right]\left[\mathbf{n}\times\left(\mathbf{v}\times\mathbf{n}\right)\right]^{a}+\frac{1}{2}\left(\mathbf{v}\cdot\mathbf{n}\right)\left[\mathbf{v}\times\left(\mathbf{n}\times\mathbf{v}\right)\right]^{a}\right\} +\mathcal{O}\left(\frac{v^{4}}{c^{4}}\right)\label{eq:sa-3}\end{eqnarray}
which is formula\eqref{eq:rel-s-n-grem} for the aberration in GREM
if we consider the case of GR where $\gamma\mathrm{_{\mathrm{PPN}}}=1$
and we take into account that $\mathbf{v}\equiv\dot{\mathbf{x}}_{\mathrm{o}}$,
and $U\equiv w\left(\mathbf{x}_{\mathrm{o}}\right)$.

Finally, the result obtained with eq.~\eqref{eq:sa-3} states that,
limited to the case of aberration and using the appropriate definitions
of the IAU recommendations, RAMOD recovers GREM at the $\left(v/c\right)^{3}$
order.

\section{\label{sec:comments}Conclusions}

This paper compares two relativistic astrometric models, GREM and
RAMOD, both suitable for modelling modern astrometric observations
at the micro-arcsecond accuracy. Their different mathematical structures
hinder a straightforward comparison and call for a more in-depth analysis
of the two models. Because of the structure of GREM, the earliest
stage of a theoretical comparison starts with the evaluation of the
aberration {}``effect'' in RAMOD. In this regard, we can evidence
the following differences in: (i) the choice of the boundary conditions,
(ii) the tools needed to define the astrometric measurements, (iii)
the attitude implementation, (iv) the definition of the proper light
direction.

Crucial is point (i). The light signal arriving at the local BCRS
along the spatial direction $l^{\alpha}=P(u)_{\beta}^{\alpha}k^{\beta}$
satisfies the RAMOD master equations, namely a set of non-linear coupled
differential equations \citep{2006ApJ...653.1552D}. Therefore the
cosines (i.e. the astrometric measurements) taken as a function of
the local line-of-sight (the \emph{physical one}), at the time of
observation ($l_{(0)}^{i}$), allow  to fix the boundary conditions
needed to solve the master equations and to determine \emph{uniquely}
the star coordinates. However, since the direction cosines are expressed
in terms of the attitude, the mathematical characterization of the
attitude frame is \emph{essential} to complete the boundary value
problem in the process of reconstructing the light trajectory. The
vector $\mathbf{n}$, i.e. the {}``aberration-free'' counterpart
of $l^{\alpha}$ in GREM, is instead used to derive the aberration
effect (in a coordinate language) and there is no need to connect
it with a RAMOD-like boundary value problem.

As for the solution of the geodesic equation, RAMOD defines a complete
procedure to derive the satellite attitude which depends as input
only on the specific terms of the metric that describes the addressed
physical problem. GREM, instead, embeds the definitions of its main
reference system (BCRS) within the metric, consequently each further
step depends on this choice. This includes all the subsequents transformations
among the reference systems which are essential to extract the GREM
observable as function of the astrometric unknowns. On the other side,
the RAMOD analytical solution for the attitude frame assures controlled
alghoritms that can be directly implemented in the solution of the
astrometric problem and guarantee its consistency with GR. In RAMOD
the direction cosines link the attitude of the satellite to the measurements,
compacting several reference frames useful to determine, as final
task, the stellar coordinates\emph{:} the BCRS (kinematically non-rotating
global reference rame), the CoMRS (a local reference frame comoving
with the satellite centre of mass), and the SRS (the attitude triad
of the satellite). The coordinate transformations between BCRS/CoMRS/SRS
come out naturally once the IAU conventions are adopted. This is inside
the conceptual framework of RAMOD, where the astrometric set-up allows
to trace back the light ray to the emitting star in a curved geometry,
and it is not natural to disentangle each single effect. Any approximation
can be applied \emph{a posteriori} where it is needed, case by case.
This explains items (ii) and (iii) and introduces item (iv). 

The direction cosines being physical quantities not depending on the
coordinates, are a powerful tool to compare the astrometric relativistic
models: their physical meaning allow us to correctly intepret the
astrometric parameters in terms of coordinate quantities. This justified
the conversion of the physical stellar proper direction of RAMOD into
its analgous Euclidean coordinate counterpart, which ultimately leads
to the derivation of a GREM-style aberration formula. Another point
arises when the observables of RAMOD have to be identified with components
of the observed $s^{i}$ of GREM. This matching is admitted only if
the origins of the boosted local BCRS tetrad in RAMOD and of the CoMRS
in GREM concide.

To what extent the process of star coordinate {}``reconstruction''
is consistent with GR\&Theory of Measurements? Solving the astrometric
problem in practice means to compile an astrometric catalogue at same
order of accuracy of the measurements. This paper shows that, already
at the level of the aberration effect, a correct treatment of physical
meaurements in terms of coordinate quantities needs particular care
in order to avoid misunderstandings in the interpretation of the quantities
which constitute the final catalogue. 

As a closing consideration, the computation of the BCRS stellar direction
in GREM needs to extract, at a second stage, the deflection terms
from the coordinate {}``aberration-free'' direction $n^{i}$ . This
problem in RAMOD is, again, embedded in the formulation of the astrometric
problem as {}``global solution'' which aims at recovering the star
coordinates by integration of the geodesic equations (treated with
an appropriate physical boundary condition and approriate reference
systems) where the deflection terms play the most fundamental role.

\begin{acknowledgments}
The authors wish to thank Prof. Fernando de Felice and Dr. Mario G.~Lattanzi
for the constant support and useful discussions. This work is supported
by the ASI grants COFIS and I/037/08/0.
\end{acknowledgments}

\bibliographystyle{apsrev}

\begin{thebibliography}{19}
\expandafter\ifx\csname natexlab\endcsname\relax\def\natexlab#1{#1}\fi
\expandafter\ifx\csname bibnamefont\endcsname\relax
  \def\bibnamefont#1{#1}\fi
\expandafter\ifx\csname bibfnamefont\endcsname\relax
  \def\bibfnamefont#1{#1}\fi
\expandafter\ifx\csname citenamefont\endcsname\relax
  \def\citenamefont#1{#1}\fi
\expandafter\ifx\csname url\endcsname\relax
  \def\url#1{\texttt{#1}}\fi
\expandafter\ifx\csname urlprefix\endcsname\relax\def\urlprefix{URL }\fi
\providecommand{\bibinfo}[2]{#2}
\providecommand{\eprint}[2][]{\url{#2}}

\bibitem[{\citenamefont{IAU}(2000)}]{2000IAU-res....B1.3}
\bibinfo{author}{\bibnamefont{IAU}}, \emph{\bibinfo{title}{{Definition of
  Barycentric Celestial Reference System and Geocentric Celestial Reference
  System}}} (\bibinfo{year}{2000}), \bibinfo{note}{iAU Resolution B1.3 adopted
  at the 24th General Assembly, Manchester, August 2000}.

\bibitem[{\citenamefont{{Turon} et~al.}(2005)\citenamefont{{Turon},
  {O'Flaherty}, and {Perryman}}}]{2005tdug.conf.....T}
\bibinfo{editor}{\bibfnamefont{C.}~\bibnamefont{{Turon}}},
  \bibinfo{editor}{\bibfnamefont{K.~S.} \bibnamefont{{O'Flaherty}}},
  \bibnamefont{and} \bibinfo{editor}{\bibfnamefont{M.~A.~C.}
  \bibnamefont{{Perryman}}}, eds., \emph{\bibinfo{title}{{The Three-Dimensional
  Universe with Gaia}}} (\bibinfo{year}{2005}).

\bibitem[{\citenamefont{{Unwin} et~al.}(2008)\citenamefont{{Unwin}, {Shao},
  {Tanner}, {Allen}, {Beichman}, {Boboltz}, {Catanzarite}, {Chaboyer},
  {Ciardi}, {Edberg} et~al.}}]{2008PASP..120...38U}
\bibinfo{author}{\bibfnamefont{S.~C.} \bibnamefont{{Unwin}}},
  \bibinfo{author}{\bibfnamefont{M.}~\bibnamefont{{Shao}}},
  \bibinfo{author}{\bibfnamefont{A.~M.} \bibnamefont{{Tanner}}},
  \bibinfo{author}{\bibfnamefont{R.~J.} \bibnamefont{{Allen}}},
  \bibinfo{author}{\bibfnamefont{C.~A.} \bibnamefont{{Beichman}}},
  \bibinfo{author}{\bibfnamefont{D.}~\bibnamefont{{Boboltz}}},
  \bibinfo{author}{\bibfnamefont{J.~H.} \bibnamefont{{Catanzarite}}},
  \bibinfo{author}{\bibfnamefont{B.~C.} \bibnamefont{{Chaboyer}}},
  \bibinfo{author}{\bibfnamefont{D.~R.} \bibnamefont{{Ciardi}}},
  \bibinfo{author}{\bibfnamefont{S.~J.} \bibnamefont{{Edberg}}},
  \bibnamefont{et~al.}, \bibinfo{journal}{Publ.\ Astron.\ Soc.\ Pac.}
  \textbf{\bibinfo{volume}{120}}, \bibinfo{pages}{38} (\bibinfo{year}{2008}),
  \eprint{0708.3953}.

\bibitem[{\citenamefont{{Bini} et~al.}(2003)\citenamefont{{Bini}, {Crosta}, and
  {de Felice}}}]{2003CQGra..20.4695B}
\bibinfo{author}{\bibfnamefont{D.}~\bibnamefont{{Bini}}},
  \bibinfo{author}{\bibfnamefont{M.~T.} \bibnamefont{{Crosta}}},
  \bibnamefont{and} \bibinfo{author}{\bibfnamefont{F.}~\bibnamefont{{de
  Felice}}}, \bibinfo{journal}{Class.\ Quantum Grav.}
  \textbf{\bibinfo{volume}{20}}, \bibinfo{pages}{4695} (\bibinfo{year}{2003}).

\bibitem[{\citenamefont{{de Felice} et~al.}(2006)\citenamefont{{de Felice},
  {Vecchiato}, {Crosta}, {Bucciarelli}, and {Lattanzi}}}]{2006ApJ...653.1552D}
\bibinfo{author}{\bibfnamefont{F.}~\bibnamefont{{de Felice}}},
  \bibinfo{author}{\bibfnamefont{A.}~\bibnamefont{{Vecchiato}}},
  \bibinfo{author}{\bibfnamefont{M.~T.} \bibnamefont{{Crosta}}},
  \bibinfo{author}{\bibfnamefont{B.}~\bibnamefont{{Bucciarelli}}},
  \bibnamefont{and} \bibinfo{author}{\bibfnamefont{M.~G.}
  \bibnamefont{{Lattanzi}}}, \bibinfo{journal}{Astrophys.\ J.}
  \textbf{\bibinfo{volume}{653}}, \bibinfo{pages}{1552} (\bibinfo{year}{2006}),
  \eprint{arXiv:astro-ph/0609073}.

\bibitem[{\citenamefont{{de Felice} and {Preti}}(2006)}]{2006CQGra..23.5467D}
\bibinfo{author}{\bibfnamefont{F.}~\bibnamefont{{de Felice}}} \bibnamefont{and}
  \bibinfo{author}{\bibfnamefont{G.}~\bibnamefont{{Preti}}},
  \bibinfo{journal}{Class.\ Quantum Grav.} \textbf{\bibinfo{volume}{23}},
  \bibinfo{pages}{5467} (\bibinfo{year}{2006}).

\bibitem[{\citenamefont{{Klioner}}(2003)}]{2003AJ....125.1580K}
\bibinfo{author}{\bibfnamefont{S.~A.} \bibnamefont{{Klioner}}},
  \bibinfo{journal}{Astron.\ J.} \textbf{\bibinfo{volume}{125}},
  \bibinfo{pages}{1580} (\bibinfo{year}{2003}).

\bibitem[{\citenamefont{{Klioner} and {Kopeikin}}(1992)}]{1992AJ....104..897K}
\bibinfo{author}{\bibfnamefont{S.~A.} \bibnamefont{{Klioner}}}
  \bibnamefont{and} \bibinfo{author}{\bibfnamefont{S.~M.}
  \bibnamefont{{Kopeikin}}}, \bibinfo{journal}{Astron.\ J.}
  \textbf{\bibinfo{volume}{104}}, \bibinfo{pages}{897} (\bibinfo{year}{1992}).

\bibitem[{\citenamefont{Lattanzi et~al.}(2006)\citenamefont{Lattanzi, Drimmel,
  Gai, and Spagna}}]{2006oato:avu}
\bibinfo{author}{\bibfnamefont{M.~G.} \bibnamefont{Lattanzi}},
  \bibinfo{author}{\bibfnamefont{R.}~\bibnamefont{Drimmel}},
  \bibinfo{author}{\bibfnamefont{M.}~\bibnamefont{Gai}}, \bibnamefont{and}
  \bibinfo{author}{\bibfnamefont{A.}~\bibnamefont{Spagna}},
  \bibinfo{type}{Tech. Rep.} (\bibinfo{year}{2006}),
  \bibinfo{note}{{GAIA-C3-TN-INAF-ML-001-2}}.

\bibitem[{\citenamefont{{Kopeikin} and {Sch{\"
  a}fer}}(1999)}]{1999PhRvD..60l4002K}
\bibinfo{author}{\bibfnamefont{S.~M.} \bibnamefont{{Kopeikin}}}
  \bibnamefont{and} \bibinfo{author}{\bibfnamefont{G.}~\bibnamefont{{Sch{\"
  a}fer}}}, \bibinfo{journal}{Phys.\ Rev.\ D} \textbf{\bibinfo{volume}{60}},
  \bibinfo{pages}{124002} (\bibinfo{year}{1999}).

\bibitem[{\citenamefont{{Kopeikin} and {Mashhoon}}(2002)}]{2002PhRvD..65f4025K}
\bibinfo{author}{\bibfnamefont{S.~M.} \bibnamefont{{Kopeikin}}}
  \bibnamefont{and}
  \bibinfo{author}{\bibfnamefont{B.}~\bibnamefont{{Mashhoon}}},
  \bibinfo{journal}{Phys.\ Rev.\ D} \textbf{\bibinfo{volume}{65}},
  \bibinfo{pages}{064025} (\bibinfo{year}{2002}).

\bibitem[{\citenamefont{{de Felice} et~al.}(2004)\citenamefont{{de Felice},
  {Crosta}, {Vecchiato}, {Lattanzi}, and {Bucciarelli}}}]{2004ApJ...607..580D}
\bibinfo{author}{\bibfnamefont{F.}~\bibnamefont{{de Felice}}},
  \bibinfo{author}{\bibfnamefont{M.~T.} \bibnamefont{{Crosta}}},
  \bibinfo{author}{\bibfnamefont{A.}~\bibnamefont{{Vecchiato}}},
  \bibinfo{author}{\bibfnamefont{M.~G.} \bibnamefont{{Lattanzi}}},
  \bibnamefont{and}
  \bibinfo{author}{\bibfnamefont{B.}~\bibnamefont{{Bucciarelli}}},
  \bibinfo{journal}{Astrophys.\ J.} \textbf{\bibinfo{volume}{607}},
  \bibinfo{pages}{580} (\bibinfo{year}{2004}).

\bibitem[{\citenamefont{de~Felice and {Clarke}}(1990)}]{1990recm.book.....D}
\bibinfo{author}{\bibfnamefont{F.}~\bibnamefont{de~Felice}} \bibnamefont{and}
  \bibinfo{author}{\bibfnamefont{C.~J.~S.} \bibnamefont{{Clarke}}},
  \emph{\bibinfo{title}{{Relativity on curved manifolds}}}
  (\bibinfo{publisher}{{Cambridge University Press}}, \bibinfo{year}{1990}).

\bibitem[{\citenamefont{{Teyssandier} and {Le
  Poncin-Lafitte}}(2006)}]{2006gr.qc....11078T}
\bibinfo{author}{\bibfnamefont{P.}~\bibnamefont{{Teyssandier}}}
  \bibnamefont{and} \bibinfo{author}{\bibfnamefont{C.}~\bibnamefont{{Le
  Poncin-Lafitte}}}, \bibinfo{journal}{ArXiv General Relativity and Quantum
  Cosmology e-prints}  (\bibinfo{year}{2006}), \eprint{gr-qc/0611078}.

\bibitem[{\citenamefont{{Crosta}}(2003)}]{2003PhDTh...MRA...C}
\bibinfo{author}{\bibfnamefont{M.~T.} \bibnamefont{{Crosta}}}, Ph.D. thesis,
  \bibinfo{school}{Universit\`a di Padova, Centro Interdipartimentale di Studi
  e Attivit\`a Spaziali (CISAS) ``G. Colombo''} (\bibinfo{year}{2003}).

\bibitem[{\citenamefont{{Misner} et~al.}(1973)\citenamefont{{Misner}, {Thorne},
  and {Wheeler}}}]{1973grav.book.....M}
\bibinfo{author}{\bibfnamefont{C.~W.} \bibnamefont{{Misner}}},
  \bibinfo{author}{\bibfnamefont{K.~S.} \bibnamefont{{Thorne}}},
  \bibnamefont{and} \bibinfo{author}{\bibfnamefont{J.~A.}
  \bibnamefont{{Wheeler}}}, \emph{\bibinfo{title}{Gravitation}}
  (\bibinfo{publisher}{San Francisco: W.H.~Freeman and Co.},
  \bibinfo{year}{1973}).

\bibitem[{\citenamefont{{Jantzen} et~al.}(1992)\citenamefont{{Jantzen},
  {Carini}, and {Bini}}}]{1992AnnPhys.215..1J}
\bibinfo{author}{\bibfnamefont{R.~T.} \bibnamefont{{Jantzen}}},
  \bibinfo{author}{\bibfnamefont{P.}~\bibnamefont{{Carini}}}, \bibnamefont{and}
  \bibinfo{author}{\bibfnamefont{D.}~\bibnamefont{{Bini}}},
  \bibinfo{journal}{Ann. Phys.} \textbf{\bibinfo{volume}{215}},
  \bibinfo{pages}{1} (\bibinfo{year}{1992}).

\bibitem[{\citenamefont{{Bastian}}(2004)}]{2004rcn.tech.ll003B}
\bibinfo{author}{\bibfnamefont{U.}~\bibnamefont{{Bastian}}},
  \bibinfo{type}{Research Note} \bibinfo{number}{GAIA-ARI-BAS-003},
  \bibinfo{institution}{GAIA livelink} (\bibinfo{year}{2004}).

\bibitem[{\citenamefont{{Klioner}}(2004)}]{2004PhRvD..69l4001K}
\bibinfo{author}{\bibfnamefont{S.~A.} \bibnamefont{{Klioner}}},
  \bibinfo{journal}{Phys.\ Rev.\ D} \textbf{\bibinfo{volume}{69}},
  \bibinfo{pages}{124001} (\bibinfo{year}{2004}).

\end{thebibliography}

\end{document}